\documentclass{article}
\usepackage{waspaa23,amsmath,graphicx,url,times}
\usepackage{color}
\usepackage{acronym}
\usepackage{svg}
\usepackage{cite}
\usepackage{flushend} 
\usepackage[citecolor=black,
            colorlinks=true,
            linkcolor=black,
            urlcolor  = black,
            pdftitle={The Effect of Spoken Language on Speech Enhancement using Self-Supervised Speech Representation Loss Functions},
            pdfauthor={George Close, Thomas Hain, and Stefan Goetze},
            pdfkeywords={Speech enhancement, self-supervised speech representations,  language, domain adaption, neural networks},
            pdfsubject={Speech enhancement, self-supervised speech representations,  language, domain adaption, neural networks}]{hyperref}
\usepackage{tabu}
\usepackage[english]{babel}
\usepackage{enumitem} 
\setlist[itemize]{leftmargin=*,  
                 labelsep=2pt,
                 nosep}  
\addto\extrasenglish{}
\addto\extrasenglish{}
\addto\extrasenglish{}
\addto\extrasenglish{}
\addto\extrasenglish{}
\addto\extrasenglish{}
\usepackage{ifthen}
\acrodef{ASR}  {Automatic Speech Recognition}
\acrodef{BLSTM}{bidirectional long short-term memory}
\acrodef{CEC}  {Clarity Enhancement Challenge}
\acrodef{CNN}  {convolutional neural network}
\acrodef{CPC}  {Clarity Prediction Challenge}
\acrodef{CPC1} {Clarity Prediction Challenge 1}
\acrodef{dBHL} {decibels hearing level}
\acrodef{DFT}  {discrete Fourier transform}
\acrodef{GAN}  {Generative Adversarial Network}
\acrodef{h2h}  {human-to-human}
\acrodef{h2m}  {human-to-machine}
\acrodef{HA}   {hearing aid}
\acrodef{HASPI}{Hearing-Aid Speech Perception Index}
\acrodef{HL}   {hearing loss}
\acrodef{HLS}  {hearing loss simulator}
\acrodef{HSR}  {human speech recognition}
\acrodef{MBSTOI}{Modified Binaural Short-Time Objective Intelligibility}
\acrodef{MOS}  {mean opinion score}
\acrodef{MMSE} {minimum mean squared error}
\acrodef{MSE}  {mean squared error}
\acrodef{NN}   {neural network}
\acrodef{OLA}  {Overlap-Add}
\acrodef{PESQ} {Perceptual Evaluation of Speech Quality}
\acrodef{RMSE} {Root Mean Square Error}
\acrodef{SE}   {speech enhancement}
\acrodef{SI}   {speech intelligibility}
\acrodef{SISDR}{scale-invariant signal-to-distortion ratio}
\acrodef{SNR}  {signal-to-noise ratio}
\acrodef{SPIN} {Speech In Noise}
\acrodef{SSSR} {self-supervised speech representation}
\acrodef{STFT} {short time Fourier transform}
\acrodef{STOI} {short-time objective intelligibility}
\acrodef{VAD}  {voice activity detector}
\acrodef{VB-D} {VoiceBank-DEMAND}
\acrodef{WER}  {Word Error Rate}
\acrodef{XLSR} {Cross-Lingual Speech Representation}

\newcommand*\rot{\rotatebox{90}}

\title{The Effect of Spoken Language on Speech Enhancement using Self-Supervised Speech Representation Loss Functions}

\name{George Close, Thomas Hain, and Stefan Goetze\thanks{This work was supported by the Centre for Doctoral Training in Speech and Language Technologies (SLT) and their Applications funded by UK Research and Innovation [grant number EP/S023062/1]. This work was also funded in part by TOSHIBA Cambridge Research Laboratory}}
\address{Speech and Hearing Group, Department of Computer Science, University of Sheffield, United Kingdom\\
}

\begin{document}

\ninept
\maketitle


\begin{abstract}
Recent work in the field of speech enhancement (SE) has involved the use of self-supervised speech representations (SSSRs) as feature transformations in loss functions. However, in prior work, very little attention has been paid to the relationship between the language of the audio used to train the self-supervised representation and that used to train the SE system. Enhancement models trained using a loss function which incorporates a self-supervised representation that shares exactly the language of the noisy data used to train the SE system show better performance than those which do not match exactly. This may lead to enhancement systems which are language specific and as such do not generalise well to unseen languages, unlike models trained using traditional spectrogram or time domain loss functions. In this work, SE models are trained and tested on a number of different languages, with self-supervised representations which themselves are trained using different language combinations and with differing network structures as loss function representations. These models are then tested across unseen languages and their performances are analysed. It is found that the training language of the self-supervised representation appears to have a minor effect on enhancement performance, the amount of training data of a particular language, however, greatly affects performance. 

\end{abstract}


\section{Introduction}
\label{sec:intro}
\Acf{SE} is a longstanding, active area of speech research given its myriad applications in downstream tasks~\cite{Loizou13SpeechEnhancementBook,MCSignalEnhancementDocloEtAl,Moritz2017,FFASRHaebUmbach}. With the change in working patterns globally due to the COVID-19 pandemic, online remote meetings have become a mainstay in the working world. \Ac{SE} systems which are able to remove various kinds of  background noise while maintaining the quality and intelligibility of speech are required features of online meeting software. Deep learning approaches utilising neural networks have shown better performance versus traditional signal-processing-based methods~\cite{fu2021metricgan,WHAMR,mimospeech,close2023PAMGAN}. Another related topic of research is the development of sophisticated ways of measuring the degradation in a noisy signal and thus the effectiveness of speech enhancement systems~\cite{PESQ,STOI,LeRoux,Avila2016Quality,close22_interspeech}. These measures (or `metrics') are often perceptually motivated, and aim in someway to behave similarly to human ratings of quality; some recent works propose neural networks which, given audio as input directly, predict human \ac{MOS}  quality ratings of the audio\cite{cauchi2019QualityLSTM,reddy2021dnsmos,MOSnet}.\\
The use of \acp{SSSR} has become increasingly popular in various speech-related tasks, including speech enhancement \cite{wav2vec,hubert,superb}. \acp{SSSR} are particularly advantageous due to their ability to predict speech content's contextual information within the input audio, thereby capturing patterns in spoken language.\\
Recent work \cite{close2023perceive,meta_phone_aware_se,perceptual_quality_phone_fort} has utilised \acp{SSSR} in loss functions for the training of neural speech enhancement systems. Loss functions which use distances between \ac{SSSR} representations have shown to be more strongly correlated with perceptual measures of quality and with human \ac{MOS} scores compared to standard \ac{STFT} spectrogram-based loss functions~\cite{close2023perceive}. However, these works focus mainly on the use of different intermediate representations from the \ac{SSSR} as features for loss functions and are not concerned with the relationship between the data used to train the \ac{SSSR} and the data used to train the enhancement system. \\
This work investigates the relationship between the spoken language in the training data of the \ac{SSSR} and the noisy speech data used to train the enhancement system. Hence, several monolingual noisy speech datasets are created, differing only in the language of the clean speech which is artificially corrupted. All other dataset parameters are kept as similar as possible. Subsequently, several speech enhancement models are trained using these datasets with \ac{SSSR}-derived loss functions where the \acp{SSSR} have been trained on different combinations of languages, as well as with standard non-\ac{SSSR} loss functions.  The performance of these models is evaluated across testsets for each language, assessing their ability to adapt to enhancing speech in languages not observed in training. \\
The remainder of this paper is structured as follows. In 
\autoref{sect:sssrs}, the \acp{SSSR} used in this work are introduced. In \autoref{sect:cv-d}, the simulation process to generate the monolingual noisy speech datasets is detailed. In \autoref{sect:exp}, an experiment involving the training of neural speech enhancement systems using the proposed datasets is carried out, and performance analysed. \autoref{sec:conclusion} concludes the paper.
\section{Self Supervised Speech Representation Models}\label{sect:sssrs}
\ac{SSSR} models are neural networks which output a \emph{context} representation of the input speech audio waveform~\cite{wav2vec}. They are trained in a self-supervised manner using large corpora of speech data, typically with the objective to recreate some masked portion of the input audio. Structurally, they consist of two main stages. The first, denoted by the operator $\mathcal{G}_\mathrm{FE}$ in the following with  subscript $_\mathrm{FE}$ standing for \emph{feature encoder}, is built from a number of $1$D convolutional layers which convert the input time-domain speech signal $s[n]$ to a two-dimensional feature representation
\begin{equation}
    \label{eq:fe_transform}
    \mathbf{S}_\mathrm{FE} = \mathcal{G}_\mathrm{FE}({s}[n]),
\end{equation} with a feature dimension $F$  (typically of size $512$) and a time dimension $T$, i.e.~the number of frames, which is dependent on the length of the input audio signal.
The second stage, denoted by $\mathcal{G}_\mathrm{OL}$, consists of a number of Transformer~\cite{transformer} layers and operates over the output of the feature encoder stage to result in the \ac{SSSR} output 
\begin{equation}
    \label{eq:ol_transform}
    \mathbf{S}_\mathrm{OL} = \mathcal{G}_\mathrm{OL}(\mathcal{G}_\mathrm{FE}(\mathbf{s}[n])).
\end{equation}
The output representation $\mathbf{S}_\mathrm{OL}$ shares the time dimension $T$ with $\mathbf{S}_\mathrm{FE}$ but has a different, usually larger feature dimension $F$. Subscript $_\mathrm{OL}$ stands for {\em output layer} of the \ac{SSSR}.\\
\subsection{SSSR Representations}\textbf{Hidden Unit BERT (HuBERT)}~\cite{hubert} is an \ac{SSSR} model; during training it makes use of a BERT~\cite{bert} inspired loss function. The output of its feature encoder $\mathcal{G}_\mathrm{FE}$ has a dimension of $F=512$, while its final layer output after $\mathcal{G}_\mathrm{OL}$ has a feature dimension of $758$. In this work, the HuBERT model used is trained on the $960$ hour Librispeech~\cite{7178964} training set and is sourced from the fairseq Github repository\footnote{\url{https://github.com/facebookresearch/fairseq}}. It is important to note that this dataset consists of English read speech only, so the model has only ever been exposed to English speech.\\ 
\textbf{Multilingual HuBERT (mHuBERT)}~\cite{lee2022textless} is a variation on HuBERT which has been trained on multilingual speech data, specifically the English, French and Spanish language parts of the VoxPopuli~\cite{voxpopuli} dataset, each containing $4.5$k hours totalling $13.5$k hours of speech. It has the same feature dimensions as HuBERT. It can be considered as a {\em middle point} between the monolingual HuBERT and the massively multilingual XLSR.\\ 
The \textbf{Cross-Lingual Speech Representation (XLSR)}~\cite{xlsr} is an \ac{SSSR} with the main distinguishing feature being that it is trained using audio containing a large number of languages. It is intended to act as a `universal' model of speech, encoding latent speech representations which are shared across languages. It is trained on $436$k hours of speech from $128$ different languages from datasets including VoxPopuli, CommonVoice and BABEL\footnote{\url{https://catalog.ldc.upenn.edu/byyear}}, with the Wav2Vec2~\cite{wav2vec} contrastive masking objective. Note that unlike the other two \acp{SSSR} used in this work, it is trained on potentially noisy data, notably CommonVoice (described below in \autoref{ssec:CommonVoiceDataset}) and BABEL which contains conversational telephone recordings. Its $\mathcal{G}_\mathrm{FE}$ representations have a feature dimension $F$ of $512$ while its $\mathcal{G}_\mathrm{OL}$ representations have an $F$ of $1024$.

\section{CommonVoice-DEMAND: A Multilingual Speech Enhancement Dataset}\label{sect:cv-d}

This section details the creation process of the proposed \linebreak\emph{ CommonVoice-DEMAND} speech enhancement dataset, which is intended to be a multilingual variation on the popular monolingual \ac{VB-D} dataset. 
\subsection{VoiceBank-DEMAND Dataset}\label{subsect:vbd}
\acl{VB-D}~\cite{ValentiniBotinhao2016InvestigatingRS} is a widely used dataset for speech enhancement neural network training. It consists of clean English read speech, artificially corrupted with environmental noise from the DEMAND\cite{demand} noise dataset, as well as two additional noise types, speech-shaped noise (SSN) and babble. Note that the babble noise was created by randomly overlapping the clean speech audio. The clean speech files vary in length from around $3$ to $10$ seconds, while the DEMAND noise recordings are all $10$ minutes long. The noisy signal $x[n]$ 
is created by adding a random part of the noise recording $v[n]$ of the same length as the clean speech $s[n]$.
\begin{equation}
    \label{eq:sim}
    x[n] = s[n] + c \cdot v[n]
\end{equation}
This scaling factor 
\begin{equation}
    \label{eq:scaling}
    c = \sqrt{\frac{P_s}{P_v \cdot 10^{\frac{\mathrm{SNR}}{10}}}}
 \end{equation}
is computed using a given target mixing \ac{SNR}, and using the ITU-T P.56 method \cite{itu-p56} for computing the active speaker power of the clean (speech) reference audio $P_s$ and of the noise $P_v$.\\
The training set consists of $11572$ pairs of clean and noisy speech, $s[n],x[n]$, from $28$ different speakers ($14$ male, $14$ female) with native British accents speaking English. The clean speech is mixed at $0$, $5$, $10$ and $15$~dB \ac{SNR} with cafeteria, car, kitchen, meeting, metro, restaurant, station, and traffic noise from DEMAND as well as babble and speech-shaped noise. The test set consists of $824$ $s[n],x[n]$ pairs from two additional speakers (one male, one female) who do not appear in the training set, mixed at $2.5$, $7.5$, $12.5$ and $17.5$~dB SNR with bus, cafe, living room, office and public square noise from DEMAND. All audio has a sample rate of $48000$~Hz and is in WAV format, but the dataset is typically down-sampled to $16000$~Hz for speech enhancement training. 

\subsection{CommonVoice Dataset}
\label{ssec:CommonVoiceDataset}
To create a multilingual dataset which is as similar as possible to the VB-D dataset described in \autoref{subsect:vbd}, speech is soured from the Mozilla CommonVoice~\cite{commonvoice:2020} dataset, which consists of recordings of read speech in 108 languages, with corresponding text prompt sentences. The recordings are crowd-sourced using the CommonVoice website\footnote{\url{https://commonvoice.mozilla.org}}. Validation that the recordings properly represent the prompt sentence is also crowd-sourced.  In addition to the audio recording and prompt sentence text, some additional metadata is sometimes available such as gender and accent of the speaker. In this work, we make use of the English, Spanish, 
and Welsh portions of the dataset. These portions contain $3209$, $2152$
and $152$ hours of audio, respectively.

\subsection{Candidate Selection}\label{subsec:reference_select}
Due to their crowd-sourced nature, the quality of the recordings in the CommonVoice dataset varies considerably. For the creation of the multilingual CommonVoice-DEMAND dataset, the aim is to select the cleanest possible audio for use as reference signals. Additionally, certain signal enhancement metrics require the input audio to have a minimum length. The process for the selection of candidate reference signals for a given language to create the CommonVoice-DEMAND datasets can be summarised as follows:\\
Firstly, only recordings which have been validated by the crowd-sourced validation process are selected. This is to ensure that the audio does contain the prompt sentence and is not a failed recording or too noisy for the speech to be intelligible.\\
Secondly, recordings of less than $2$ seconds length and those which contain a single-word utterance are excluded since it was found that some speech enhancement metrics have difficulties assessing such recordings.\\
Finally, the quality of the remaining audio recordings is assessed. A \ac{VAD} is used to segregate frames of length $L$ of the signal $x[n]$ into disjoint sets $\mathcal{A}$ and $\mathcal{B}$, for which the signal fulfils either the hypothesis that speech is present $\mathcal{H}_1$, or that speech is absent $\mathcal{H}_0$, respectively. An \ac{SNR} estimate is obtained by
\begin{equation}
\widehat{\mathrm{SNR}}(x[n]) = 10\log_{10}\left(\frac{\frac{1}{|\mathcal{A}|}\sum_{\ell\in \mathcal{A}} \sum_{n=0}^{K-1} x^2[\ell K+n]}{\frac{1}{|\mathcal{B}|}\sum_{\ell\in\mathcal{B}} \sum_{n=0}^{K-1} x^2[\ell K+n]}\right)
\end{equation}
for frame index $\ell$ with  $|\mathcal{A}|$ and $|\mathcal{B}|$ denoting the respective cardinalities of the sets $\mathcal{A}$ and $\mathcal{B}$.
For simplicity, the Google WebRTC-VAD\footnote{\url{https://github.com/wiseman/py-webrtcvad}} is used. A minimum threshold of $50$~dB estimated \ac{SNR} by the above formulation is used to select the candidate recordings. Note that while this is a somewhat crude estimator in that it does not account for the noise which is present in the speech-active frames in $\mathcal{A}$, it was found empirically that this approach works sufficiently well to select high-quality recordings containing little to no background noise, with a low computation overhead. For each language, $20000$ of such candidate recordings are selected in this way.  
During this process, the candidate recordings are converted from the MP3 format to the WAV format. 
\subsection{Dataset Creation}
In order to ensure that the proposed CommonVoice-DEMAND datasets are comparable to the original VoiceBank-DEMAND dataset, the log files describing VoiceBank-DEMAND are used. These consist of two lists (one for the training set, one for the test set) 
for clean audio file $s[n]$, the name of the noise file for $v[n]$, from which a random section of the same length as $s[n]$  is obtained and the desired mixing \ac{SNR} value. The process for the creation of the CommonVoice-DEMAND datasets is as follows:
\begin{itemize}
    \item A candidate CommonVoice recording is selected which is closest to the length in seconds to the clean audio recording from the original VoiceBank dataset. This candidate recording is either padded with zero values or truncated such that it is the same number of samples exactly as the original VoiceBank recording.
    \item The mixing process as described in \autoref{subsect:vbd} is carried out, using the selected CommonVoice recording as $s[n]$, and creating a corresponding noisy version $x[n]$ according to (\ref{eq:sim}) and (\ref{eq:scaling}). Since CommomVoice data has a sampling rate of $32$~kHz, the resulting sampling rate of CommonVoice-DEMAND is also $32$~kHz and thus lower than VoiceBank-DEMAND's $48$~kHz. Note, that for \ac{SE}, data is usually anyhow downsampled to $16$~kHz.
    \item The selected CommonVoice recording is then removed from the list of available candidate recordings, ensuring that uniqueness for each $s[n],x[n]$ pair in the resultant dataset. 
\end{itemize}
The goal of this process is to create a dataset which is as similar as possible to the original VoiceBank-DEMAND dataset, but using clean speech with a different language or source. These datasets differ from the original VoiceBank-DEMAND in the (usually) greater number of speakers in the training and test sets and the exact portion of the noise file from which $v[n]$ is created due to the random sampling of the noise file. The created CommonVoice-DEMAND data ensures reproducibility by fixed seeds in the random number generator.
The code for this process is available on GitHub\footnote{\url{https://github.com/leto19/CommonVoice-DEMAND}} to allow for the recreation and reuse of the generated corpora. 
\section{Speeech Enhancement Experiments}\label{sect:exp}
\subsection{Experiment Setup}
Masking-based speech enhancement networks are trained using \ac{SSSR} derived loss functions using the proposed CommonVoice-DEMAND datasets.
The SpeechBrain~\cite{speechbrain} toolkit is used to facilitate the training and testing of the models. The models are trained for $50$ epochs with the Adam~\cite{kingma2017adam} optimiser, with a learning rate of $0.001$. At test time, the model with the highest \ac{PESQ}~\cite{PESQ} score on the validation set is loaded and evaluated. \\
The model used for all training consists of two \ac{BLSTM} layers and two linear layers. The first linear layer has LeakyReLU activation and the second has Sigmoid activation. The model takes a spectrogram of the noisy audio $x[n]$ as input and produces a \emph{mask} which is multiplied with the input spectrogram to generate an enhanced spectrogram. Using the original noisy phase of $x[n]$, the enhanced time domain signal $\hat{s}[n]$ is produced via the overlap-add resynthesis method. Despite its simplicity, this model's training is stable, efficient, and has demonstrated state-of-the-art performance~\cite{fu2021metricgan,close2022}.
\subsection{Datasets}
CommonVoice-DEMAND training and test sets were generated using the process described in \autoref{sect:cv-d} for English and Spanish. These languages were chosen as they match languages used to train the mHuBERT model, and the CommonVoice corpus component for each is sufficiently large. A testset for Welsh was also created as a language which was not used to train HuBERT or mHuBERT but which was used as one of XLSRs $128$ languages. The CommonVoice-DEMAND datasets have the same size training and test sets as the original VoiceBank-DEMAND with $11572$ and $824$ $s[n],x[n]$ pairs, respectively. A validation set of size $770$ is created from each CommonVoice-DEMAND training set. All audio is at $16$~kHz sample rate.

\subsection{SSSR Signal Enhancement Loss Function}
The \ac{SSSR} loss function as defined in \cite{close2023perceive} is used, which is based on the \ac{MSE} distance between the output feature encoder representations of the enhanced signal $\hat{s}[n]$ and the reference signal $s[n]$. The \ac{SSSR} loss function is given by
\begin{equation}
\label{eq:fe_loss}
L_{\mathrm{FE}}(\mathbf{S}_\mathrm{FE},\mathbf{\hat{S}}_\mathrm{FE}) = \frac{1}{TF}\sum_{t,f} (\mathbf{S}_{\mathrm{FE}}[t,f] - \mathbf{\hat{S}}_{\mathrm{FE}}[t,f])^2,
\end{equation}
where $\mathbf{S}_\mathrm{FE}$ and $\mathbf{\hat{S}}_\mathrm{FE}$ are the feature encoder output representations of $s[n]$ and $\hat{s}[n]$, respectively. $F$ and $T$ denote the feature and time dimensions of the representations, with $F$ being $512$ for all models used in this work and $T$ depending on the length in samples of the time domain audio. Models are trained using HuBERT, mHuBERT, and XLSR to obtain the representations in (\ref{eq:fe_loss}). In addition, the spectrogram \ac{MSE} loss, \ac{STOI}~\cite{STOI} loss, and \ac{SISDR}~\cite{LeRoux} loss are used as baselines. These baselines are popular loss functions for speech enhancement training and are language-independent. Spectrograms are created using a \ac{STFT} with a FFT length of $512$, a window length of $32$~ms, a hop length of $16$~ms, and a hamming window.

\subsection{Result}
\begin{table}[!t]
\small
\centering
\caption{Performance of models trained on CommonVoice-DEMAND \emph{English}; tested on English, Spanish and Welsh testsets.}
\begin{tabular}{rlllll}
\textbf{Model}         & \multicolumn{1}{l}{\textbf{PESQ}} & \multicolumn{1}{l}{\textbf{STOI}} & \multicolumn{1}{l}{\textbf{CSIG}} & \multicolumn{1}{l}{\textbf{CBAK}} & \multicolumn{1}{l}
{\textbf{COVL}} \\ \hline
\textit{Noisy}         & \textit{2.19}                     & \textit{0.95}                     & \textit{3.27}                     & \textit{2.40}                     & \textit{2.67}                     \\
Spec Loss              & 2.64   & 0.96   & 3.66   & 2.64   & 3.14   \\
STOI Loss              & 2.46   & \textbf{0.96}                     & 3.47   & 2.35   & 2.93   \\
SISDR Loss             & 2.73   & 0.96   & 3.45   & 2.74   & 3.07   \\
\rot{\rlap{\textbf{English}}}
HuBERT $L_{\mathrm{FE}}$    & 2.75   & 0.95   & \textbf{3.78}                     & 2.71   & \textbf{3.25}                     \\
mHuBERT $L_{\mathrm{FE}}$   & \textbf{2.79}                     & 0.96   & 3.70   & 2.76   & {3.23}                     \\
XLSR $L_{\mathrm{FE}}$      & 2.48   & 0.93   & 3.30   & \textbf{2.92}                     & 2.86   \\
                       \\ \hline
\textit{Noisy}         & \textit{2.12}                     & \textit{0.95}                     & \textit{2.98}                     & \textit{2.23}                     & \textit{2.46}                     \\
Spec Loss              & 2.57   & 0.96   & 3.37   & 2.62   & 2.94   \\
STOI Loss              & 2.39   & \textbf{0.96}                     & 3.19   & 2.31   & 2.73   \\
SISDR Loss             & 2.68   & 0.96   & 3.15   & 2.74   & 2.88   \\
\rot{\rlap{Spanish}}

HuBERT $L_{\mathrm{FE}}$    & 2.72   & 0.95   & \textbf{3.57}                     & 2.71   & \textbf{3.11}                     \\
mHuBERT $L_{\mathrm{FE}}$   & \textbf{2.75}                     & 0.96   & {3.48}                     & 2.75   & {3.08}                     \\
XLSR $L_{\mathrm{FE}}$      & 2.51   & 0.93   & 2.93   & \textbf{2.81}                     & 2.65   \\
\textbf{}              & \multicolumn{1}{l}{\textbf{}}     & \multicolumn{1}{l}{\textbf{}}     & \multicolumn{1}{l}{\textbf{}}     & \multicolumn{1}{l}{\textbf{}}     & \multicolumn{1}{l}{\textbf{}}     \\\hline
\textit{Noisy}       & \textit{2.12}                     & \textit{0.96}                     & \textit{3.06}                     & \textit{2.18}                     & \textit{2.50}                     \\
Spec Loss            & 2.61  & 0.96  & 3.39  & 2.56  & 2.97  \\
STOI Loss            & 2.45  & \textbf{0.97}                     & 3.33  & 2.27  & 2.83  \\
SISDR Loss           & 2.72  & 0.97  & 3.15  & 2.69  & 2.89  \\
\rot{\rlap{Welsh}}

HuBERT $L_{\mathrm{FE}}$  & 2.71  & 0.96  & \textbf{3.56}                     & 2.61  & 3.09  \\
mHuBERT $L_{\mathrm{FE}}$ & \textbf{2.78}                     & 0.96  & 3.48  & 2.68  & \textbf{3.09}                     \\
XLSR $L_{\mathrm{FE}}$    & 2.44  & 0.93  & 2.93  & \textbf{2.73}                     & 2.62                  
\end{tabular}
\label{tab:english-results}
\end{table}

Tables~\ref{tab:english-results} and \ref{tab:spanish-results} display results on CommonVoice-DEMAND testsets for models trained on English and Spanish, respectively. \ac{PESQ}~\cite{PESQ}, \ac{STOI}~\cite{STOI}, and the components of the Composite~\cite{composite} intrusive metrics are reported, where higher values are better. The scores for the best performing model on each testset are highlighted in bold.  The models perform better on the respective testset matching their language of training; the drop in performance on the non-matching testsets is consistent across all the models trained, including the baseline systems.  Performance on the proposed English CommonVoice-DEMAND dataset is similar to that of models trained on the original VoiceBank-DEMAND in \cite{close2023perceive}. The best performing models, in terms of CBAK score are  those trained with XLSR $L_{\mathrm{FE}}$ loss, except for one case. This is again consistent with the findings in \cite{close2023perceive}. HuBERT and mHuBERT perform similarly, with mHuBERT slightly outperforming in most cases. \\
\begin{table}[!t]
\small
\centering
    \caption{Performance of models trained on CommonVoice-DEMAND \emph{Spanish}; tested on English, Spanish and Welsh testsets.}
\begin{tabular}{rlllll}
\textbf{Model}       & {\textbf{PESQ}} & \multicolumn{1}{l}{\textbf{STOI}} & {\textbf{CSIG}} & {\textbf{CBAK}} & {\textbf{COVL}} \\ \hline
\textit{Noisy}       & \textit{2.19}                     & \textit{0.95}                     & \textit{3.27}                     & \textit{2.40}                     & \textit{2.67}                     \\
Spec Loss  & 2.50   & 0.95   & 3.54   & 2.59   & 3.00   \\
STOI Loss  & 2.37   & \textbf{0.96}                     & 3.32   & 2.35   & 2.80   \\
SISDR Loss  & 2.44   & 0.95   & 3.25   & 2.58   & 2.82   \\
\rot{\rlap{English}}
HuBERT $L_{\mathrm{FE}}$   & 2.60   & 0.95   & 3.58                     & 2.62   & 3.06               \\
mHuBERT $L_{\mathrm{FE}}$  & \textbf{2.70}                     & 0.95   & \textbf{3.62}                     & 2.69   & \textbf{3.14}                     \\
XLSR $L_{\mathrm{FE}}$     & 2.44   & 0.93   & 3.13   & \textbf{2.75}                     & 2.72   \\
                     & \multicolumn{1}{l}{}              & \multicolumn{1}{l}{}              & \multicolumn{1}{l}{}              & \multicolumn{1}{l}{}              & \multicolumn{1}{l}{}              \\\hline
\textit{Noisy}       & \textit{2.12}                     & \textit{0.95}                     & \textit{2.98}                     & \textit{2.23}                     & \textit{2.46}                     \\
Spec Loss            & 2.64   & 0.96   & 3.50   & 2.68   & 3.04   \\
STOI Loss            & 2.40   & \textbf{0.96}                     & 3.11   & 2.36   & 2.70   \\
SISDR Loss           & 2.61   & 0.96   & 3.18   & 2.72   & 2.86   \\
\rot{\rlap{\textbf{Spanish}}}
HuBERT $L_{\mathrm{FE}}$   & 2.81   & 0.96   & \textbf{3.75}                     & 2.76   & 3.25                     \\
mHuBERT $L_{\mathrm{FE}}$  & \textbf{2.89}                     & 0.96   & 3.73   & 2.81   & \textbf{3.29}                     \\
XLSR $L_{\mathrm{FE}}$     & 2.63   & 0.95   & 3.06   & \textbf{2.83}                     & 2.77   \\
                     & \multicolumn{1}{l}{}              & \multicolumn{1}{l}{}              & \multicolumn{1}{l}{}              & \multicolumn{1}{l}{}              & \multicolumn{1}{l}{}              \\\hline
\textit{Noisy}       & \textit{2.12}                     & \textit{0.96}                     & \textit{3.06}                     & \textit{2.18}                     & \textit{2.50}                     \\
Spec Loss            & 2.63  & 0.96  & 3.49  & 2.59  & 3.03  \\
STOI Loss            & 2.42  & \textbf{0.97}                     & 3.22  & 2.29  & 2.77  \\
SISDR Loss           & 2.60  & 0.97  & 3.18  & 2.63  & 2.85  \\
\rot{\rlap{Welsh}}
HuBERT $L_{\mathrm{FE}}$  & 2.72  & 0.96  & 3.62  & 2.62  & 3.13  \\
mHuBERT $L_{\mathrm{FE}}$ & \textbf{2.83}                     & 0.96  & \textbf{3.64}                     & \textbf{2.69}                     & \textbf{3.21}                     \\
XLSR $L_{\mathrm{FE}}$    & 2.52  & 0.94  & 2.94  & 2.64  & 2.65 
\end{tabular}
\label{tab:spanish-results}
\end{table}
Interestingly, all \ac{SSSR} loss function models trained using Spanish audio perform better on the Welsh testset than those trained on the English audio. This is despite the fact that Welsh is lexically more similar to English than Spanish \cite{lex_sim}.\\
The quantity of data used for training the SSSR appears to be more important than language, as mHuBERT is trained with more English audio than HuBERT, however XLSR, trained with the most English speech data, performs worse. To further investigate this, an additional model was trained utilising the WavLM Base+~\cite{wavLM} \ac{SSSR}, training and testing on the English CommonVoice-DEMAND dataset. WavLM Base+ has a  parameter count comparable to HuBERT and mHuBERT and is trained with a similar objective and but with an additional speech denoising task. It is trained on $96$k hours of English only audio. These results are shown in \autoref{tab:wavlm-results}; the new model performs similarly in terms of PESQ score but somewhat better than all others in terms of CSIG. \\
Overall, these results suggest that the BERT style training objective HuBERT, mHuBERT and WavLM might make them better suited as loss function feature representations when signal quality is the main concern as shown by the high \ac{PESQ} and CSIG scores while the contrastive feature encoding masking objective of XLSR makes it more suitable if the objective of the enhancement is background noise reduction at the cost of additional speech distortion as the higher CBAK scores of the XLSR models demonstrates. 
\begin{table}[!ht]
\centering
\small
\caption{Performance of $L_{FE}$ Loss models trained on CommonVoice-DEMAND English and tested on English testset.}
\begin{tabular}{rlllll}
\textbf{Model}       & \multicolumn{1}{l}{\textbf{PESQ}} & \multicolumn{1}{l}{\textbf{STOI}} & \multicolumn{1}{l}{\textbf{CSIG}} & \multicolumn{1}{l}{\textbf{CBAK}} & \multicolumn{1}{l}{\textbf{COVL}} \\ \hline
\textit{Noisy}       & \textit{2.19}                     & \textit{0.95}                     & \textit{3.27}                     & \textit{2.40}                     & \textit{2.67}                     \\
HuBERT $L_{\mathrm{FE}}$  & 2.75  & 0.95  & 3.78  & 2.71  & {3.25}                     \\
mHuBERT $L_{\mathrm{FE}}$ & \textbf{2.79}                     & 0.96  & 3.70  & 2.76  & 3.23  \\
XLSR $L_{\mathrm{FE}}$    & 2.48  & 0.93  & 3.30  & \textbf{2.92}                     & 2.86  \\
WavLM $L_{\mathrm{FE}}$   & 2.76  & \textbf{0.96}  & \textbf{3.84}                     & 2.71  & \textbf{3.28}                    
\end{tabular}

\label{tab:wavlm-results}
\end{table}
\section{Conclusion}
\label{sec:conclusion}
In this work, a system to create noisy speech datasets for a number of languages are proposed. These noisy speech datasets are used to train and test neural speech enhancement models using \ac{SSSR}  based loss functions. It is found that the language of the audio used to train the representations has a minimal impact on their performance when used in this manner, and that training objective and amount of training data has a greater effect. However, questions remain as the degree that these factors effect the performance of these loss functions; future work will aim to investigate this.

\bibliographystyle{IEEEtran}
 {\footnotesize
\bibliography{refs23}
}
\end{document}